\documentclass[12pt]{article}
\usepackage{changes}
\usepackage{amsmath}
\usepackage{graphicx,psfrag,epsf}
\usepackage{enumerate}
\usepackage{natbib}
\usepackage{url} 
\usepackage{multirow}
\usepackage{booktabs}
\usepackage{caption}
\usepackage{amssymb}
\usepackage{float}
\usepackage{authblk}
\graphicspath{{figures/}}     
\newcommand{\blind}{1}

\newcommand{\mbf}{\mathbf}

\newcommand{\T}{\top}

\newtheorem{theorem}{Theorem}[section]

\newtheorem{exam}{Example}[section]

\begin{document}

\def\spacingset#1{\renewcommand{\baselinestretch}%
{#1}\small\normalsize} \spacingset{1}


\if1\blind
{
  \title{\bf Optimal Penalized Function-on-Function Regression under a Reproducing Kernel Hilbert Space Framework}
\author[1]{Xiaoxiao Sun \thanks{The authors are grateful to Dr. Xiaoyu Zhang who kindly provided
the histone regulation data and pertinent explanations of the experiments.
Du's research was supported by U.S. National Science Foundation
under grant DMS-1620945.  Sun and Ma's research was supported
by U.S. National Science Foundation under grants DMS-1440037 and
DMS-1438957 and by U.S. National Institute of Health under grants 1R01GM122080-01.
Wang's research was supported by U.S. National Science Foundation under grant DMS-1613060.}}
\author[2]{Pang Du}
\author[3]{Xiao Wang}
\author[1]{Ping Ma}
\affil[1]{Department of Statistics, University of Georgia}
\affil[2]{Department of Statistics, Virginia Tech}
\affil[3]{Department of Statistics, Purdue University}
\date{}

  \maketitle
} \fi

\if0\blind
{
  \bigskip
  \bigskip
  \bigskip
  \begin{center}
    {\LARGE\bf Title}
\end{center}
  \medskip
} \fi

\bigskip
\begin{abstract}
Many scientific studies collect data where the response and predictor variables are both functions of time, location, or some other covariate. Understanding the relationship between these functional variables is a common goal in these studies. Motivated from two real-life examples, we present in this paper a function-on-function regression model that can be used to analyze such kind of functional data. Our estimator of the 2D coefficient function is the optimizer of a form of penalized least squares where the penalty enforces a certain level of smoothness on the estimator. Our first result is the Representer Theorem which states that the exact optimizer of the penalized least squares actually resides in a data-adaptive finite dimensional subspace although the optimization problem is defined on a function space of infinite dimensions. This theorem then allows us an easy incorporation of the Gaussian quadrature into the optimization of the penalized least squares, which can be carried out through standard numerical procedures. We also show that our estimator achieves the minimax convergence rate in mean prediction under the framework of function-on-function regression. Extensive simulation studies demonstrate the numerical advantages of our method over the existing ones, where a sparse functional data extension is also introduced. The proposed method is then applied to our motivating examples of the benchmark Canadian weather data and a histone regulation study.
\end{abstract}

\noindent%
\vfill

\newpage
\spacingset{1.45} 
\section{Introduction}
\label{sec:intro}

Functional data have attracted much attention in the past decades \citep{ramsay2005functional}. Most of the existing literature has only considered the regression models of a scalar response against one or more functional predictors, possibly with some scalar predictors as well. Some of them considered a reproducing kernel Hilbert space framework. For example,  \cite{yuancai:10}  provided a thorough theoretical analysis of the penalized functional linear regression model with a scalar response. The paper laid the foundation for several theoretical developments including the representer theorem and minimax convergence rates for prediction and estimation for penalized functional linear regression models. In a follow-up, \cite{cai2012minimax} showed that the minimax rate of convergence for the excess prediction risk is determined by both the covariance kernel and the reproducing kernel. Then they designed a data-driven roughness regularization predictor that can achieve the optimal convergence rate adaptively without the knowledge of the covariance kernel. \cite{duwang:13} extended the work of \cite{yuancai:10} to the setting of a generalized functional linear model, where the scalar response comes from an exponential family distribution.

In contrast to these functional linear regression models with a scalar response, the model with a functional response $Y(t)$ over a functional predictor $X(s)$ has only been scarcely investigated \citep{yao2005functional, ramsay2005functional}. Such data with functional responses and predictors are abundant in practice.  We shall now present two motivating examples.

\begin{figure}[H]
\centering
\includegraphics[scale=0.45]{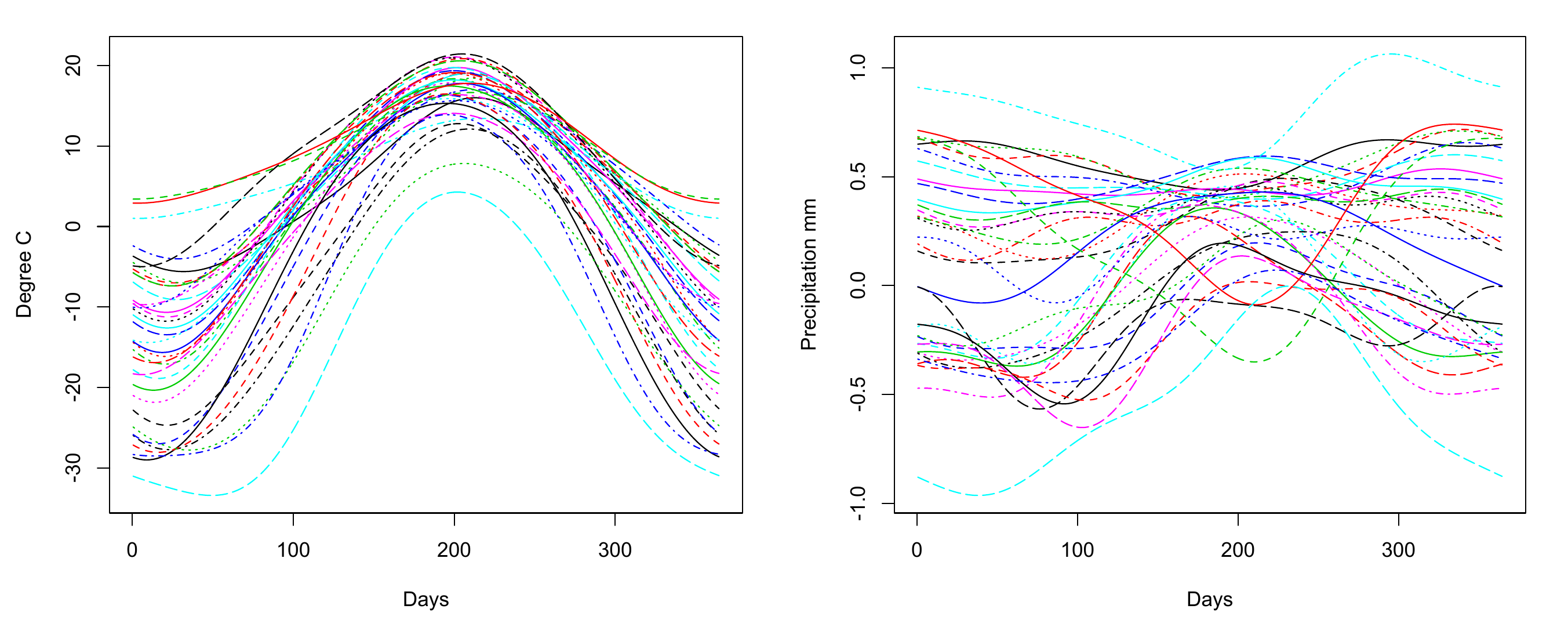}
\caption{ Smoothed trajectories of temperature (Celsius) in left panel and the $\log$ (base 10) of daily precipitation (Millimetre) in right panel. The x-axis labels in both panels represent 365 days.}
\label{fig:canadian}
\end{figure}

\begin{exam}
\label{canadian}
\textbf{Canadian Weather Data} \newline
Daily temperature and precipitation at 35 different locations in Canada averaged over 1960 to 1994 were collected (Figure \ref{fig:canadian}). The main interest is to use the daily temperature profile to predict the daily precipitation profile for a location in Canada.
\end{exam}

\begin{figure}[H]
\centering
\includegraphics[scale=0.45]{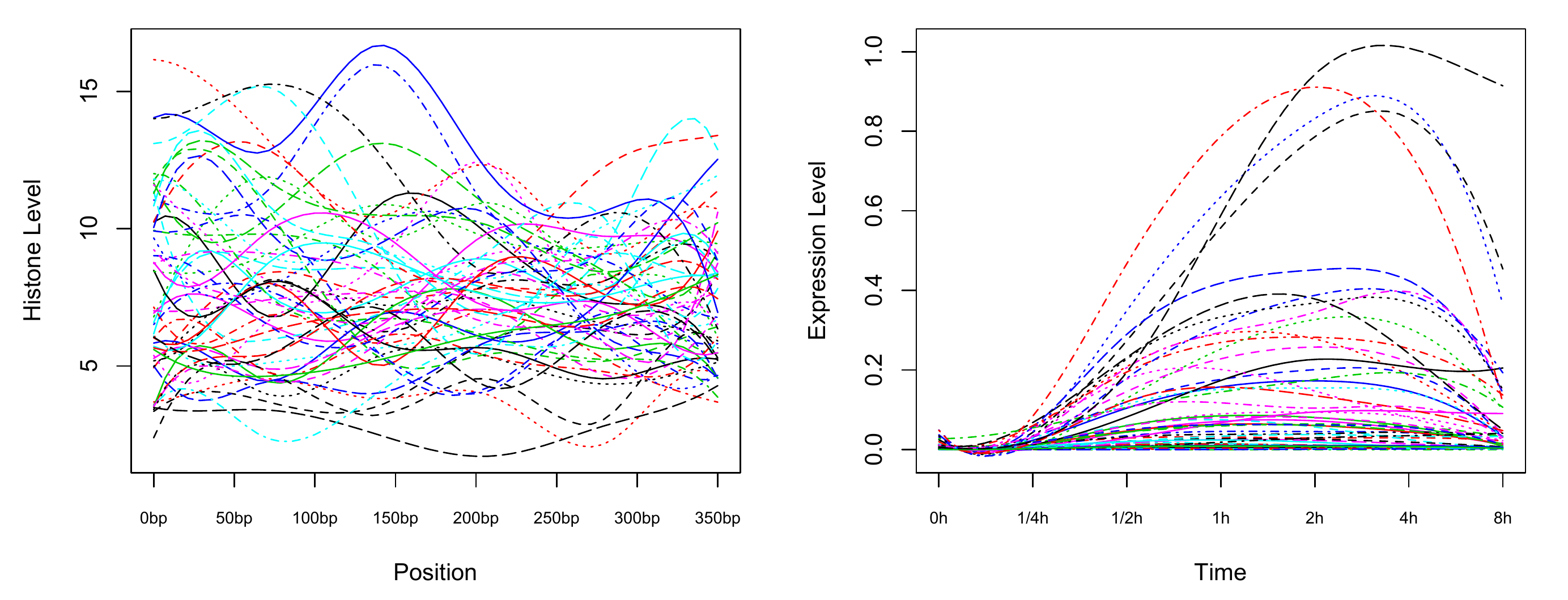}
\caption{Smoothed trajectories of normalized histone levels in ChIP-seq experiments in left panel and the normalized expression levels in RNA-seq experiments in right panel. The x-axis label in the left panel stands for the region of 350 bp. The x-axis label in the right panel represents seven time points within 24 hours.}\label{fig:hisrna}
\end{figure}

\begin{exam}
\label{histone}
\textbf{Histone Regulation Data} \newline
Extensive researches have been shown that histone variants, i.e. histones with structural changes compared to their primary sequence, play an important role in the regulation of chromatin metabolism and gene activity \citep{ausio2006histone}. An ultra-high throughput time course experiment was conducted to study the regulation mechanism during heat stress in \textit{Arabidopsis thaliana}.
The genome-wide histone variant distribution was measured by ChIP sequencing (ChIP-seq) \citep{johnson2007genome} experiments. We computed histone levels over 350 base pairs (bp) on genomes from the ChIP-seq data, see left panel in Figure \ref{fig:hisrna}. 
The RNA sequencing (RNA-seq) \citep{wang2009rna} experiments measured the expression levels over seven time points within 24 hours, see right panel in Figure \ref{fig:hisrna}.  Of primary interest is to study the regulation mechanism between gene expression levels over time domain and histone levels over spatial domain.

\end{exam}

Motivated by the examples, we now present the statistical model.
Let $\big\{ (X(s), Y(t)): s\in I_x, t\in I_y \big\}$ be two random processes defined respectively on $I_x, I_y\subseteq \mathbb R$.
Suppose $n$ independent copies of $\big(X, Y\big)$ are observed: $\big(X_i(s), Y_i(t)\big)$,  $i=1, \ldots, n$.
The functional linear regression model of interest is
\begin{equation}\label{equ:model1}
Y_i(t) = \alpha(t) + \int_{I_x} \beta\big(t, s\big)X_i(s) ds + \epsilon_i(t), ~~~~ t\in I_y,
\end{equation}
where $\alpha(\cdot): I_y\rightarrow \mathbb R$ is the intercept function,
$\beta(\cdot, ~\cdot): I_y\times I_x\rightarrow \mathbb R$ is a bivariate coefficient function,
and $\epsilon_i(t)$, independent of $X_i(s)$, are i.i.d. random error functions with $\mathbb E\epsilon_i(t) =0$ and
$\mathbb E \|\epsilon_i(t)\|_2^2< \infty$. Here $\|\cdot\|_2$ denotes the $L_2$-norm. In Example \ref{canadian}, $Y_i(t)$ and $X_i(t)$ represent the daily precipitation and temperature at station $i$. In Example \ref{histone}, the expression levels of gene $i$ over seven time points, $Y_i(t)$, from RNA-seq is used as the functional response.  The histone levels of gene $i$ over 350 base pairs (bp), $X_i(s)$, from ChIP-seq is used as the functional predictor. 

At a first look, model \eqref{equ:model1} might give the (wrong) impression of being an easy extension from the model with a scalar response, with the latter obtained from \eqref{equ:model1} by removing all the $t$ notation. However, the coefficient function in the scalar response case is univariate and thus can be easily estimated by most off-the-shelf smoothing methods. When extended to estimating a bivariate coefficient function $\beta(t,s)$ in \eqref{equ:model1}, many of these smoothing methods may encounter major numerical and/or theoretical difficulties. This partly explains the much less abundance of research in this direction.

Some exceptions though are reviewed below. \cite{cuevas:02} considered a fixed design case, a different setting from \eqref{equ:model1} with $Y_i(t)$ and $X_i(s)$ represented and
analyzed as sequences. Nonetheless they provided many motivating applications in neuroscience, signal transmission, pharmacology, and chemometrics, where \eqref{equ:model1} can apply.
The historical functional linear model in \cite{hist:03} was among the first to study regression of a response functional variable over
a predictor functional variable, or more precisely, the history of the predictor function.
 \cite{ferraty:11} proposed a simple extension of the classical Nadaraya-Watson estimator to the functional case and derived its convergence
rates. They provided no numerical results on the empirical performance of their kernel estimator.
\cite{ben:15} extended ridge regression to the functional setting.
However, their estimation relied on an empirical estimate of the covariance process of predictor functions. Theoretically sound as it is, this covariance process estimate is generally not reliable in practice. Consequently, their coefficient surface estimates suffered as shown in their simulation plots.
\cite{meyer2015bayesian} proposed a Bayesian function-on-function regression model for multi-level functional data,
where the basis expansions of functional parameters were regularized by basis-space prior distributions and a random effect function was introduced to incorporate the with-subject correlation between functional observations.

A popular approach has been the functional principal component analysis (FPCA) as in \cite{yao2005functional} and \cite{crambes:13}.
The approach starts with a basis representation of $\beta(t,s)$ in terms of the eigenfunctions
in the Karhunen-Lo\`{e}ve expansions of $Y(t)$ and $X(s)$.
Since this representation has infinitely many terms, it is truncated at certain point to obtain an estimable basis expansion of $\beta(t,s)$.
\cite{yao2005functional} studied a general data setting where $Y(t)$ and $X(s)$ are only sparsely observed at some random points.
They derived the consistency and proposed asymptotic point-wise confidence bands for predicting response trajectories.
\cite{crambes:13} furthered the theoretical investigation of the FPCA approach by providing a minimax optimal rates in terms of the mean
square prediction error. However, the FPCA approach has a couple of critical drawbacks.
Firstly, $\beta(t,s)$ is a statistical quantity unrelated to $Y(t)$ or $X(s)$. Hence the leading
eigenfunctions in the {\it truncated} Karhunen-Lo\`{e}ve expansions of $Y(t)$ and $X(s)$ may not be an effective basis for representing $\beta(t,s)$. See, e.g., \cite{cai2012minimax} and \cite{duwang:13} for some scalar-response
examples where the FPCA approach breaks down when the aforementioned situation happens.
Secondly, the truncation point is integer-valued and thus only has a discrete control on the model complexity.
This puts it at disadvantage against the roughness penalty regularization approach, which offers a continuous control via a positive and real-valued smoothing parameter \cite[Chapter 5]{ramsay2005functional}.

In this paper, we consider a penalized function-on-function regression approach to estimating the bivariate coefficient function $\beta(t,s)$. There have been a few recent developments in the direction of penalized function-on-function regression. \cite{lian2015minimax} studied the convergence rates of the function-on-function regression model under a reproducing kernel Hilbert space framework.  Although his model resembled model \eqref{equ:model1}, he developed everything with the variable $t$ fixed and did not enforce any regularization on the $t$ direction. Firstly, this lack of $t$-regularization can be problematic since this leaves the noisy errors on the $t$ direction completely uncontrolled and can result in an $\beta(s,t)$ estimate that is very rough on the $t$ direction. Secondly, this simplification of fixing $t$ essentially reduces the problem to a functional linear model with a scalar response and thus makes all the results in \cite{yuancai:10} directly transferrable even without calling on any new proofs. The R package \texttt{fda} maintained by Ramsay et al. has implemented a version of penalized B-spline estimation of $\beta(t,s)$ with a fixed smoothing parameter.
\cite{ivanescu2015penalized} considered a penalized function-on-function regression model where the coefficient functions were represented by expansions into some basis system such as tensor cubic B-splines. Quadratic penalties on the expansion coefficients were used to control the smoothness of the estimates. This work provided a nice multiple-predictor-function extension to the function-on-function regression model in the \texttt{fda} package. \cite{scheipl2016ident} studied the identifiability issue in these penalized function-on-function regression models.
However, this penalized B-spline approach has several well-known drawbacks. First, it is difficult to show any theoretical optimality such as the minimax risk of mean prediction in \cite{cai2012minimax}. So its theoretical soundness is hard to justify. Moreover, the B-spline expansion is only an approximate solution to the optimization of the penalized least squares score. Hence the penalized B-spline estimate is not numerically optimal from the beginning either. These drawbacks can have negative impacts on the numerical performance as we shall see from the simulation results in Section \ref{sec:verify}.

The penalized function-on-function regression method proposed in this paper obtains its estimator of $\beta(t,s)$ through the minimization of penalized least squares on a reproducing kernel Hilbert space that is naturally associated with the roughness penalty. Such a natural formulation through a reproducing kernel Hilbert space offers several advantages comparing with the existing penalized function-on-function regression methods. Firstly,
it allows us to establish a Representer Theorem which states that, although the optimization of the penalized least squares is defined on an infinite dimensional function space, its solution actually resides in a data-adaptive finite dimensional subspace. This result guarantees an exact solution when the optimization is carried out on this finite dimensional subspace. This result itself is a nontrivial generalization of the Representer Theorems in the scenarios of nonparametric smooth regression model \citep{wahba:90} and the penalized functional regression model with a scalar response \citep{yuancai:10}.
Based on the Representer Theorem, we propose an estimation algorithm which uses penalized least squares and Gaussian quadrature with the Gauss-Legendre rule to estimate the bivariate coefficient function. The smoothing parameter is selected by the generalized cross validation (GCV) method. Secondly, the reproducing kernel Hilbert space framework allows us to show that our estimator has the optimal rate of mean prediction since it achieves the minimax convergence rate in terms of the excess risk. This generalizes the results in \cite{cai2012minimax} and \cite{duwang:13} for functional linear regression with a scalar response to the functional response scenario. 
In the numerical study, we have also considered the problem with sparsely sampled data. Particularly, we introduce  an extra pre-smoothing step before applying the proposed penalized functional regression model.  The pre-smoothing step implements the principal-component-analysis-through-expectation (PACE) method in \cite{yao2005functional2}.
Our extensive simulation studies demonstrate the numerical advantages of our method over the existing ones. In summary, our method has the following distinguishing features: (i) it makes no structural dependence assumptions of $\beta(t,s)$ over the predictor and response processes; (ii) the Representer Theorem guarantees an exact solution instead of an approximation to the optimization of the penalized score; (iii) benefited from the Representer Theorem, we develop a numerically reliable algorithm that has sound performance in simulations; (iv) we show theoretically the estimator achieves the optimal minimax convergence rate in mean prediction.

The rest of the paper is organized as follows. In Section \ref{sec:meth}, we first derive a Representer Theorem showing that the solution of the minimization of penalized least squares can be found in a finite-dimension subspace. In addition, an easily implementable estimation algorithm is considered in the Section \ref{sec:meth}. In Section \ref{sec:opt}, we prove that our method has the optimal rate of mean prediction. Numerical experiments are reported in Section \ref{sec:verify}, where we compare our method with the functional linear regressions in \citep{ramsay2005functional,yao2005functional} in terms of prediction accuracy. Two real data examples, the Canadian weather data, and the histone regulation data are analyzed in Section \ref{sec:real}. Discussion in Section \ref{sec:conc} concludes the paper. Proofs of the theorems are collected in Supplementary Material.

\section{Penalized Functional Linear Regression Method}
\label{sec:meth}

We first introduce a simplification to model \eqref{equ:model1}.  Since model \eqref{equ:model1} implies that
$$Y_i(t) - \mathbb E Y_i(t) = \int_{I_x}\beta(t, s) \{X_i(s) - \mathbb E X_i(s)\}ds + \epsilon_i(t),~~~~ t\in I_y,$$
we may, for simplicity, only consider $X$ and $Y$ to be centered, i.e., $\mathbb E X = \mathbb E Y =0$. Thus, the functional linear regression model takes the form of
\begin{equation}\label{equ:model2}
Y_i(t) = \int_{I_x} \beta\big(t, s\big)X_i(s) ds + \epsilon_i(t), ~~~~ t\in I_y.
\end{equation}

\subsection{The Representer Theorem}

Assume that the unknown $\beta$ resides in a reproducing kernel Hilbert space ${\cal H}(K)$ with the reproducing kernel $K: I\times I\rightarrow \mathbb R$, where $I = I_y\times I_x$.
The estimate $\hat \beta_{n}$ can be obtained by minimizing the following penalized least squares functional
\begin{equation}\label{equ:obj}
{1\over n}\sum_{i=1}^n \int_{I_y}\Big\{Y_i(t) - \int_{I_x}\beta(t, s) X_i(s)ds \Big\}^2dt + \lambda J(\beta)
\end{equation}
with respect to $\beta \in{\cal H}(K)$, where the sum of integrated squared errors represents the goodness-of-fit, $J$ is a roughness penalty on $\beta$, and $\lambda>0$ is the smoothing parameter balancing the trade-off.
We now establish a Representer Theorem stating that $\hat{\beta}_n$ actually resides in a finite dimensional subspace of $\mathcal{H}(K)$. This result generalizes Theorem 1 in \cite{yuancai:10} and facilitates the computation by reducing an infinite dimensional optimization problem to a finite dimensional one.

Note that the penalty functional $J$ is a squared semi-norm on ${\cal H}(K)$. Its null space
$
{\cal H}_0 = \{\beta\in {\cal H}(K): J(\beta) = 0\}
$
is a finite-dimensional linear subspace of ${\cal H}(K)$. Denote by ${\cal H}_1$ its orthogonal complement in ${\cal H}(K)$ such that ${\cal H}(K) = {\cal H}_0\oplus {\cal H}_1$. For any $\beta\in {\cal H}(K)$, there exists a unique decomposition $\beta = \beta_0+\beta_1$ where $\beta_0 \in {\cal H}_0$ and $\beta_1 \in {\cal H}_1$. Let $K_0(\cdot, \cdot)$ and $K_1(\cdot, \cdot)$ be the corresponding reproducing kernels of ${\cal H}_0$ and ${\cal H}_1$. Then $K_0$ and $K_1$ are both nonnegative definite operators on $L_2$, and $K = K_0+K_1$. In fact the penalty term $J(\beta) =\|\beta\|_{K_1}^2 =\|
\beta_1\|_{K_1}^2$.
By the theory of reproducing kernel Hilbert spaces, 
$\mathcal{H}(K)$ has a tensor product decomposition
$\mathcal{H}(K)=\mathcal{H}_y(K_y)\otimes\mathcal{H}_x(K_x)$. Here $\mathcal{H}_y(K_y)$ is the reproducing kernel Hilbert space with a reproducing kernel $K_y: I_y\times I_y\rightarrow \mathbb{R}$, and $\mathcal{H}_x(K_x)$ is the reproducing kernel Hilbert space with a reproducing kernel $K_x: I_x\times I_x\rightarrow \mathbb{R}$. For the reproducing kernels, we have $K(t,s)=K_y(t)K_x(s)$. Note that the functions in $\mathcal{H}_y(K_y)$ and $\mathcal{H}_x(K_x)$ are univariate and defined respectively on $I_y$ and $I_x$.
Similar to the decomposition of $\mathcal{H}$ and $K$, we have the tensor sum decompositions of the marginal subspaces
$\mathcal{H}_y(K_y)=\mathcal{H}_{0y}\oplus\mathcal{H}_{1y}$ and
$\mathcal{H}_x(K_x)=\mathcal{H}_{0x}\oplus\mathcal{H}_{1x}$, and the orthogonal decompositions of the marginal reproducing kernels
$K_y=K_{0y}+K_{1y}$ and $K_x=K_{0x}+K_{1x}$. Here $K_*$ is a reproducing kernel on $\mathcal{H}_*$ with $*$ running through the index set
$\{0y,1y,0x,1x\}$.

Upon piecing the marginal decomposition parts back to the tensor product space, we obtain
$\mathcal{H}_0=\mathcal{H}_{0y}\otimes\mathcal{H}_{0x}$ and
$\mathcal{H}_1=(\mathcal{H}_{0y}\otimes\mathcal{H}_{1x})\oplus(\mathcal{H}_{1y}\otimes\mathcal{H}_{0x})\oplus
(\mathcal{H}_{1y}\otimes\mathcal{H}_{1x})$. Correspondingly, the reproducing kernels satisfy that
\begin{align*}
K_0((t_1,s_1),(t_2,s_2))&=K_{0y}(t_1,t_2)K_{0x}(s_1,s_2),\\
K_1((t_1,s_1),(t_2,s_2))&=K_{0y}(t_1,t_2)K_{1x}(s_1,s_2)+K_{1y}(t_1,t_2)K_{0x}(s_1,s_2)+K_{1y}(t_1,t_2)K_{1x}(s_1,s_2).
\end{align*}
Let $N_y=\text{dim}(\mathcal{H}_{0y})$ and $N_x=\text{dim}(\mathcal{H}_{0x})$.
Denote by $\{\psi_{k,y}: k=1,\ldots,N_y\}$ and $\{\psi_{l,x}: l=1,\ldots,N_x\}$ respectively the basis functions of $\mathcal{H}_{0y}$ and $\mathcal{H}_{0x}$. With some abuse of notation, define $(K_{1y} g)(\cdot)=\int_{I_y}K_{1y}(\cdot,t)g(t)dt$ and
$(K_{1x} f)(\cdot)=\int_{I_x}K_{1x}(\cdot,s)f(s)ds$.
Now we can state the Representer Theorem as follows with its proof collected in the Supplementary Material.

\begin{theorem}
\label{th:rep}

Let $\hat{\beta}_n$ be the minimizer of \eqref{equ:obj} in $\mathcal{H}(K)$. Then $\hat{\beta}_n$ resides in the subspace of functions of the form
\begin{align}\nonumber
\beta(t,s)&=\Big\{\sum_{k=1}^{N_y}d_{k,\beta_y}\psi_{k,y}(t)+\sum_{i=1}^n c_{i,\beta_y}(K_{1y} Y_i)(t)\Big\}
\Big\{\sum_{l=1}^{N_x}d_{l,\beta_x}\psi_{l,x}(s)+\sum_{j=1}^n c_{j,\beta_x}(K_{1x} X_j)(s)\Big\}\\
&=\Big\{d_{\beta_y}^{\T}\psi_{y}(t)+c_{\beta_y}^{\T}(K_{1y} Y)(t)\Big\}
\Big\{d_{\beta_x}^{\T}\psi_{x}(s)+c_{\beta_x}^{\T}(K_{1x} X)(s)\Big\},
\label{equ:rep}
\end{align}
where $d_{\beta_y}=(d_{1,\beta_y},\ldots,d_{N_y,\beta_y})^{\T}$, $c_{\beta_y}=(c_{1,\beta_y},\ldots,c_{n,\beta_y})^{\T}$,
$d_{\beta_x}=(d_{1,\beta_x},\ldots,d_{N_x,\beta_x})^{\T}$ and $c_{\beta_x}=(c_{1,\beta_x},\ldots,c_{n,\beta_x})^{\T}$ are some coefficient vectors,
and $\psi_x, \psi_y, K_{1y}Y$ and $K_{1x}X$ are vectors of functions.
\end{theorem}

For the purpose of illustration, we give a detailed example below.
\begin{exam}\label{ex1}
Consider the case of tensor product cubic splines with $I_y=I_x=[0,1]$. The marginal spaces
$\mathcal{H}_y(K_y)=\mathcal{H}_x(K_x)=\{g: \int_0^1 (g'')^2<\infty\}$ with the inner product
\[\langle f,g\rangle_{\mathcal{H}_y}=\Big(\int_0^1f\int_0^1 g+\int_0^1f'\int_0^1 g'\Big)+\int_0^1 f''g''dt.\]
The marginal space $\mathcal{H}_y(K_y)$ can be further decomposed into the tensor sum of
$\mathcal{H}_{0y}=\{g: g''=0\}$ and $\mathcal{H}_{1y}=\{g: \int_0^1 g=\int_0^1 g'=0, \int_0^1 (g'')^2<\infty\}$.
The reproducing kernel $K_y$ is the orthogonal sum of $K_{0y}(t_1,t_2)=1+r_1(t_1)r_1(t_2)$ and
$K_{1y}(t_1,t_2)=r_2(t_1)r_2(t_2)-r_4(|t_1-t_2|)$, where $r_\nu(t)=B_\nu(t)/\nu!$ is a scaled version of the Bernoulli polynomial $B_\nu$.
The space $\mathcal{H}_{0y}$ has a dimension of $N_y=2$ and a set of basis functions $\{1, r_1(t)\}$.

The function space $\mathcal{H}(K)$ is defined as
$\mathcal{H}(K)=\{\beta: J(\beta)<\infty\}$ with the reproducing kernel $K(t,s)=K_y(t)K_x(s)$ and the penalty functional
\begin{multline*}
J(\beta)=\int_0^1\Big[\Big\{\int_0^1 \frac{\partial^2}{\partial s^2}\beta(t,s)dt\Big\}^2
+\Big\{\int_0^1 \frac{\partial^3}{\partial t\partial s^2}\beta(t,s)dt\Big\}^2 \Big]ds\\
+\int_0^1\Big[\Big\{\int_0^1 \frac{\partial^2}{\partial t^2}\beta(t,s)ds\Big\}^2
+\Big\{\int_0^1 \frac{\partial^3}{\partial t^2\partial s}\beta(t,s)ds\Big\}^2 \Big]dt
+\int_0^1\int_0^1\Big\{\frac{\partial^4}{\partial t^2\partial s^2}\beta(t,s)\Big\}^2dtds
\end{multline*}
We have $\mathcal{H}(K)=\mathcal{H}_y(K_y)\otimes\mathcal{H}_x(K_x)$ and $K=K_yK_x$; see, e.g., Chapter 2 of \cite{gu:13}.
\end{exam}
\subsection{Estimation Algorithm}

To introduce the computational algorithm, we first need some simplification of notation. Let $N=N_yN_x$ and $L=n(N_y+N_x+n)$. We rewrite the functions spanning the subspace in Theorem \ref{th:rep} as $\psi_1(t,s) = \psi_{1,y}(t)\psi_{1,x}(s)$, $\cdots$, $\psi_N(t,s) = \psi_{N_y,y}(t)\psi_{N_x,x}(s)$ and $\xi_1(t,s) = \psi_{1,y}(t)(K_{1x}X_1)(s)$, $\cdots$, $\xi_L(t,s) = (K_{1y}Y_n)(t)(K_{1x}X_n)(s)$. Thus a function in this subspace has the form $\beta(t,s) = \mbf{d}^T\psi(t,s) + \mbf{c}^T\xi(t,s)$ for some coefficient vectors $\mbf{d}$, $\mbf{c}$ and vectors of functions $\psi(t,s)$, $\xi(t,s)$. To solve (\ref{equ:obj}), we choose Gaussian quadrature with the Gauss-Legendre rule to calculate the integrals. Consider the Gaussian quadrature evaluation of an integral on $I_y$ with knots $\{t_1,\cdots, t_T\}$ and weights $\{\alpha_1,\cdots, \alpha_T\}$ such that $\int_{I_y}f(t)dt=\sum_{j=1}^T\alpha_jf(t_j)$. Let $W$ be the diagonal matrix with $\alpha_1, \cdots, \alpha_T$ repeating $n$ times on the diagonal. Then the estimation of $\beta$ in \eqref{equ:obj} reduces to the minimization of
\begin{equation} \label{eq:obj-weighted}
(Y_w-S_w\mbf{d}-R_w\mbf{c})^T(Y_w-S_w\mbf{d}-R_w\mbf{c}) + n\lambda \mbf{c}^TQ\mbf{c}
\end{equation}
with respect to $\mbf{d}$ and $\mbf{c}$, where  $Y_w=W^{1/2}Y$ with $Y=(Y_1(t_1),\ldots,Y_1(t_T),\ldots,Y_n(t_1),\ldots,Y_n(t_T))^{\T}$,
$S_w=W^{1/2}S$ with $S$ being an $nT\times N$ matrix with the $((i-1)T+j,\nu)$th entry $\int_{I_x}\psi_\nu(t_j,s)X_i(s)ds$,
$R_w=W^{1/2}R$ with $R$ being an $nT\times L$ matrix with the $((i-1)T+j,k)$th entry $\int_{I_x}\xi_k(t_j,s)X_i(s)ds$, and $Q$ is a $L\times L$ matrix with the $(i,j)$th entry $\langle\xi_i,\xi_j\rangle_{\mathcal{H}_1}$.
Let $Q_x = \left[\int_0^1 \int_0^1 X_i(u)K(u,v)X_j(v)dudv \right]_{i,j=1}^n$,
$Q_y = \left[\int_0^1 \int_0^1 Y_i(u)K(u,v)Y_j(v)dudv \right]_{i,j=1}^n$, and
$Q_{xy} = Q_x \otimes Q_y$, we have $Q=\text{diag}(Q_x, Q_x, Q_y, Q_y, Q_{xy})$.

We then utilize standard numerical linear algebra procudures such as the Cholesky decomposition with pivoting and forward and back substitutions, to calculate $\mbf{c}$ and $\mbf{d}$ in \eqref{eq:obj-weighted} \cite[Section~3.5]{gu:13}. To choose the smoothing parameter $\lambda$ in \eqref{eq:obj-weighted}, a modified Generalized Cross-Validation (GCV) score \citep{wahba1979smoothing},
\begin{equation} \label{cross-validate-one}
V(\lambda) = \frac{(nT)^{-1}Y_w^T(I-A(\lambda))^2Y_w}{\{(nT)^{-1}tr(I-\alpha A(\lambda))\}^2}
\end{equation}
is implemented, where $\alpha > 1$ is a fudge factor curbing undersmoothing \citep{kim2004smoothing} and $A(\lambda)$ is the smoothing matrix bridging the prediction $\hat{Y}_w$ and the observation $Y_w$ as
$\hat Y_w = A(\lambda) Y_w$, similar to the hat matrix in a general linear model.

\section{Optimal Mean Prediction Risk}
\label{sec:opt}
We are interested in the estimation of coefficient function $\beta$ and mean prediction, that is, to recover the functional $\eta_{\beta}(X, \cdot) = \int_{I_x} \beta(\cdot, s)X(s)ds$
based on the training sample $(X_i, Y_i)$, $i=1,\ldots, n$. Let $\hat \beta_n(t, s)$ be an estimate of $\beta(t, s)$.
Suppose $(X_{n+1}, Y_{n+1})$ is a new observation that has the same distribution as and is also independent of $(X_i, Y_i)$, $i=1,\ldots, n$. Then the prediction accuracy can be naturally measured by the excess risk
\begin{align*}
&{\mathfrak R}_n(\hat \beta_n)\\
 =& \int_{I_y}\left[\mathbb E^*\Big\{Y_{n+1}(t) - \int_{I_x} \hat\beta_n(t, s)X_{n+1}(s)ds\Big\}^2 - \mathbb E^*\Big\{Y_{n+1}(t) - \int_{I_x} \beta(t, s) X_{n+1}(s)ds\Big\}^2\right]dt \\
=& \int_{I_y}\mathbb E^*\Big\{\eta_{\hat\beta_n}(X_{n+1}, t) - \eta_{\beta}(X_{n+1}, t)\Big\}^2 dt
\end{align*}
where $\mathbb E^*$ represents the expectation taken over $(X_{n+1}, Y_{n+1})$ only. We shall study the convergence rate of ${\mathfrak R}_n$ as the sample size $n$ increases.

This section collects two theorems whose combination indicates that our estimator achieves the optimal minimax convergence rate in mean prediction.
We first establish the minimax lower bound for the convergence rate of the excess risk ${\mathfrak R}_n$.
There is a one-to-one relationship between $K$ and ${\cal H}(K)$ which is a linear functional space endowed with an inner product $\langle\cdot, \cdot\rangle_{{\cal H}(K)}$ such that
$$\beta(t,s) = \Big\langle K\big((t, s), \cdot\big), \beta\Big\rangle_{{\cal H}(K)}, ~~~ \mbox{for any } \beta\in {\cal H}(K).$$
The kernel $K$ can also be treated as an integral operator such that
$$K(\beta)(\cdot) = \Big\langle K\big((t,s), \cdot\big), \beta  \Big\rangle_{L_2} = \int\int_{I} K((t,s), \cdot) \beta(t, s)dtds.$$
It follows from the spectral theorem that there exist a set of orthonormal eigenfunctions $\{\zeta_k: k\ge 1\}$ and  a sequence of eigenvalues $\kappa_1\ge \kappa_2\ge \cdots>0$ such that
$$K((t_1,s_1), (t_2,s_2)) = \sum_{k=1}^\infty \kappa_k \zeta_k(t_1,s_1)\zeta_k(t_2,s_2), ~~~~K(\zeta_k) = \kappa_k \zeta_k, ~~k=1,2,\ldots.$$
Denote
$K^{1/2}\big((t_1,s_1), (t_2,s_2))\big) = \sum_{k=1}^\infty \kappa_k^{1/2} \zeta_k(t_1,s_1)\zeta_k(t_2,s_2).$
Let $C(t, s) = \mathrm{cov}\big(X(t), X(s)\big)$ be the covariance kernel of $X$. Define a new kernel $\Pi$ such that
\begin{equation}\label{equ:kerP}
\Pi\big((t_1, s_1), (t_2, s_2)\big) = \int\int\int_{I_x\times I_x\times I_y}   K^{1/2}\big((t_1,s_1), (z, u)\big)  ~C(u, v)~  K^{1/2}\big((t_2,s_2), (z, v)\big) dudvdz.
\end{equation}
Let $\rho_1\ge\rho_2\ge \cdots>0$ be the eigenvalues of $\Pi$ and $\{\phi_j: j\ge 1\}$ be the corresponding eigenfunctions. Therefore,
$$\Pi\big((t_1, s_1), (t_2, s_2)\big)= \sum_{k=1}^\infty \rho_k \phi_k(t_1, s_1)\phi_k(t_2, s_2), ~~~\forall ~(t_1, s_1),(t_2, s_2) \in I_y\times I_x.$$

\begin{theorem} \label{thm:upper}
Assume that for any $\beta\in L_2([0,1]^2)$
\begin{equation}\label{equ:condition}
\int\mathbb E\Big( \int \beta(t, s)X(s)dt \Big)^4 dt \le c\int \Big( \mathbb E\Big( \int \beta(t, s)X(s)ds \Big)^2  \Big)^2dt
\end{equation}
for a positive constant $c$. Suppose that the eigenvalues $\{\rho_k: k\ge 1\}$ of the kernel $\Pi$ in (\ref{equ:kerP}) satisfy $\rho_k \asymp k^{-2r}$ for some constant $0<r<\infty$. Then,
\begin{equation}
\lim_{A\rightarrow \infty}\lim_{n\rightarrow\infty}\sup_{\beta \in {\cal H}(K)}\mathbb P\Big\{ \mathfrak{R}_n \ge A n^{-{2r\over 2r+1}}  \Big\} = 0,
\end{equation}
when $\lambda$ is of order $n^{-2r/(2r+1)}$.
\end{theorem}

Theorem \ref{thm:upper} indicates that the convergence rate is determined by the decay rate of the eigenvalues of this new operator $\Pi$, which is jointly determined by both reproducing kernel $K$ and the covariance kernel $C$ as well as the alignment between $K$ and $C$ in a complicated way. This result has not been reported in the literature before. A close and related result is from \cite{yuancai:10} who studied an optimal prediction risk for functional linear models, where the optimal rate depends on the decay  rate of the eigenvalues of $K^{1/2}CK^{1/2}$.
It is interesting to see, on the other hand, whether the convergence rate of $\hat\beta_n$ in Theorem \ref{thm:upper} is
optimal. In the following, we derive a minimax lower bound for the risk.

\begin{theorem}\label{thm:low}
Let $r$ be as in Theorem~\ref{thm:upper}. Then the excess prediction risk satisfies
\begin{equation}
\lim_{c\rightarrow 0}\lim_{n\rightarrow\infty} \inf_{\tilde \eta} \sup_{\beta \in {\cal H}(K)} \mathbb P\Big( {\mathfrak R}_n \ge c n^{-{2r\over 2r+1}}  \Big) =1,
\end{equation}
where the infimum is taken over all possible predictors $\tilde \eta$ based on $\{(X_i, Y_i): i=1, \ldots, n\}$.
\end{theorem}

Theorem \ref{thm:low} shows that the minimax lower bound of the convergence rate for the prediction risk is $n^{-{2r/ 2r+1}}$, which is determined by $r$ and the decay rate of the eigenvalues of $\Pi$.
We have shown that this rate is achieved by  our penalized estimator, and therefore our estimator is rate-optimal.

\section{Numerical Experiments}
\label{sec:verify}

We compared the proposed optimal penalized function-on-function regression (OPFFR) method with existing function-on-function linear regression models under two different designs. In a dense design, each curve was densely sampled at regularly-spaced common time points. We compared the OPFFR with two existing models. In a sparse design, each curve was irregularly and sparsely sampled at possibly different time points. We extended the OPFFR to this design by adding an extra pre-smoothing step and compared it with the FPCA model. In the first model \citep{ramsay2005functional} for comparison, the coefficient function is estimated by penalizing its B-spline basis function expansion. This approach does not have the optimal mean prediction property and partially implemented in the \texttt{fda} package of R (\texttt{linmod} function) for the case of a fixed smoothing parameter. We shall add a search on the grid $10^{(-2\,:\,0.4\,:\,2)}$  for smoothing parameter selection to their implementation and denote this augmented approach by FDA. The coefficient function is represented in terms of 10 basis functions each for the $t$ and $s$ directions. 
The second model for comparison was the functional principal component analysis (hence denoted by FPCA) approach proposed by \cite{yao2005functional}. The coefficient function is represented in terms of the leading functional principal components. This is implemented in the MatLab package \texttt{PACE} (\texttt{FPCreg} function) maintained by the UC-Davis research group. The Akaike information criterion (AIC) and fraction of variance explained (FVE) criterion were used to select the number of principal components for predictor and response respectively. The cutoff value for FVE was $0.9$. The `regular' parameter was set to 2 for the dense design and 0 for the sparse design. No binning was performed.

\subsection{Simulation Study}
\subsubsection{Dense Design}

We simulated data according to model (\ref{equ:model2}) with three scenarios.
\begin{itemize}
\item
{Scenario 1}:
The predictor functions are
$X_i(s) = \sum_{k=1}^{50} (-1)^{(k+1)}k^{-1}Z_{ik}\vartheta_1(s,k)$,
where $Z_{ik}$ is from the uniform distribution $U(-\sqrt{3}, \sqrt{3})$, and $\vartheta_1(s, k) =1$ if $k=1$ and $\sqrt{2}\cos((k-1)\pi s)$ otherwise.
The coefficient function $\beta(t,s)= e^{-(t+s)}$ is the exponential function of $t$ and $s$.

\item
{Scenario 2}:
The predictor functions $X_i(s)$ are the same as those in Scenario 1 and the coefficient function $\beta(t,s) = 4\sum_{k=1}^{50} (-1)^{(k+1)}k^{-2}\vartheta_1(t,k)\vartheta_1(s,k)$.

\item
{Scenario 3}:
The predictor functions $X_i(s)$ are generated as
$X_i(s) = \sum_{k=1}^{3} (-1)^{(k+1)}k^{-1}Z_{ik}\vartheta_2(s,k)$,
where $\vartheta_2(s, k)=1$ if $k=3$ and $\sqrt{2}\cos(k\pi s)$ otherwise.
The coefficient function $\beta(t,s)= 4\sum_{k=1}^{3} (-1)^{(k+1)}k^{-2}\vartheta_2(t,k)\vartheta_2(s,k)$.
\end{itemize}

For each simulation scenario, we generated $n=30$ samples, each with 20 time points on the interval $(0,1)$. The random errors $\epsilon(t)$ were from a normal distribution with a constant variance $\sigma^2$. The value of $\sigma$ was adjusted to deliver three levels of signal-to-noise ratio (SNR$=0.5$, 5, and 10) in each scenario.
To assess the mean prediction accuracy, we generated an additional $n^*=30$ predictor curves $\tilde{X}$ and computed the mean integrated squared error $\text{MISE} =1/n^*\sum_{i=1}^{n^*} \int_{0}^{1}(\eta_{\hat \beta}(\tilde{X}_i, t) -\eta_{\beta}(\tilde{X}_i, t))^2dt$, where $\hat \beta$ was the estimator obtained from the training data.
We had 100 runs for each combination of scenario and SNR.

\begin{figure}[h]
\centering
\includegraphics[scale=0.45]{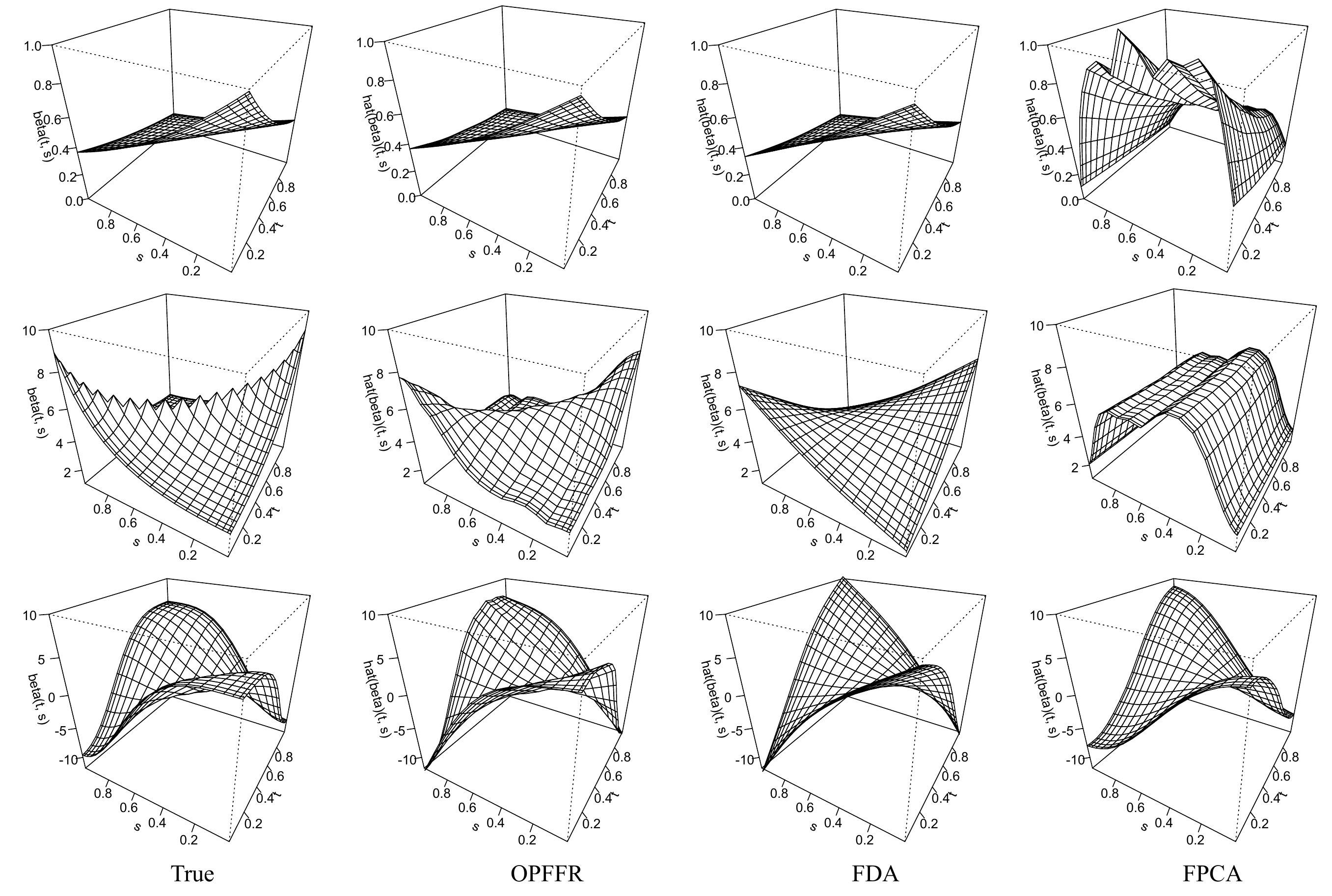}
\caption{Perspective plots of the true $\beta(t,s)$ in three scenarios, and their respective estimates by the OPFFR, FDA, and FPCA methods when SNR$=10$.}
\label{fig:beta}
\end{figure}

We applied the OPFFR, FDA and FPCA methods to the simulated data sets.
Figure \ref{fig:beta} displayed the perspective plots of the true coefficient functions in the three scenarios as well as their respective estimates for a single run with SNR$=10$.
In the first two scenarios, both OPFFR and FDA did a decent job in recovering the true coefficient function although the FDA estimates were slightly oversmoothed. In both scenarios the FPCA estimates clearly suffered since the true coefficient function could not be effectively represented by the eigen-functions of the predictor processes.

\begin{figure}[H]
\centering
\includegraphics[scale=0.65]{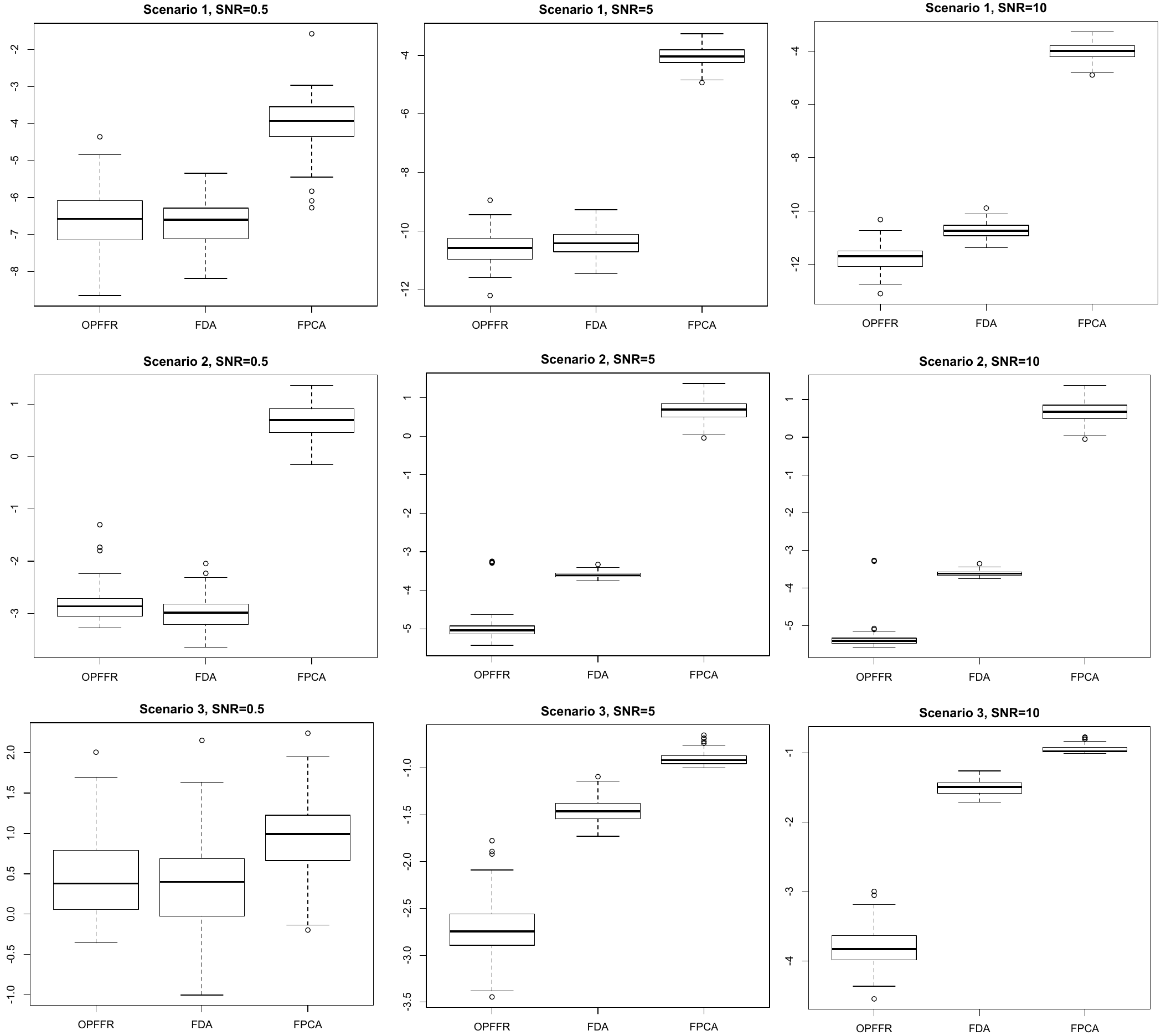}
\caption{ Boxplots of $\log_2(\text{MISE})$ for three scenarios under three signal-to-noise ratios (SNR$=0.5$, 5, 10), based on 100 simulation runs. OPFFR is the proposed approach.}
\label{fig:simu-re1}
\end{figure}

Figure \ref{fig:simu-re1} 
gave the summary reports of performances in terms of MISEs based on 100 runs. When the signal to noise ratio is low, the OPFFR and FDA approaches had comparable performances. But when the signal to noise ratio increases, OPFFR showed clear advantage against FDA. The FPCA method failed to deliver competitive performance against the other two methods in all the settings due to its restrictive requirement of the effective representation of the coefficient function.

\subsubsection{Sparse Design}
In this section, we compared the performance of the proposed OPFFR method and the FPCA method regarding  prediction error on sparsely, irregularly, and noisily observed functional data.  
	To extend our method to sparsely and noisily observed data, we first applied the principal-component-analysis-through-conditional-expectation (PACE) method in \cite{yao2005functional2} to the sparse functional data. Then we obtained a dense version of functional data by computing the PACE-fitted response and predictor functions at 50 selected time points for each curve.  We applied the OPFFR method to these densely generated data and called this sparse extension to the OPFFR by the OPFFR-S method. The original OPFFR method, FPCA and OPFFR-S methods were all applied to the simulated data for comparison.
	
We first generated $n=200$ samples for both response and predictor functions in Scenario 3, each with 50 time points on interval $(0,1)$. To obtain different sparsity levels, we then randomly chose 5, 10 and 15 time points from the 50 ones for each curve independently.  Normally distributed random errors were added to functional response and predictor with the SNR set to 10 in generating each pair of noisy response and predictor. The mean integrated squared error (MISE) was calculated based on additional $n^*=50$ predictor curves without random noises.

\begin{figure}[h]
	\centering
	\includegraphics[scale=0.5]{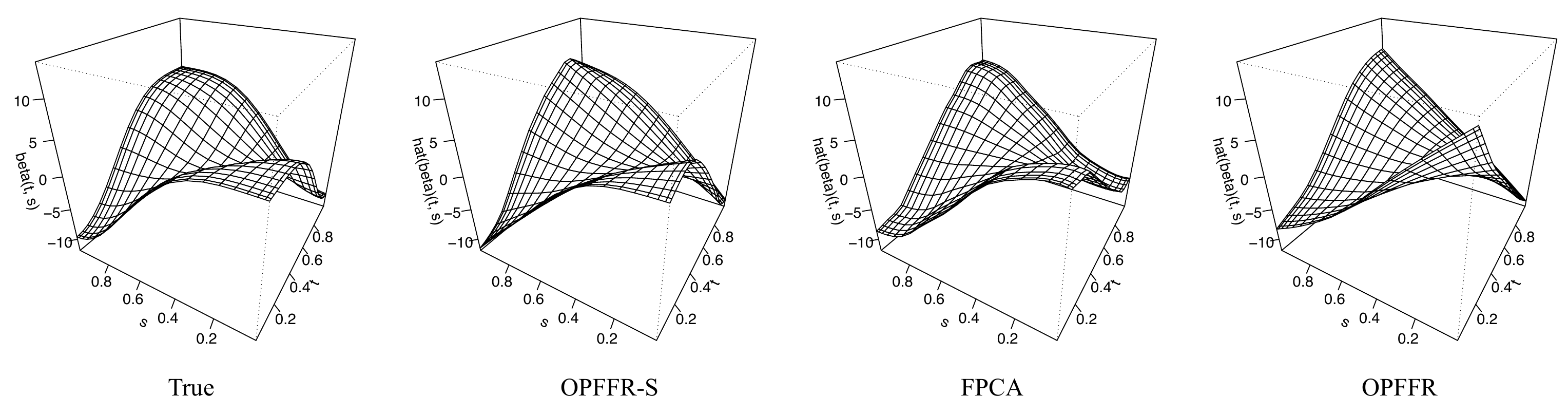}
	\caption{Perspective plots of the true $\beta(t,s)$ in the sparse scenario, and their respective estimates by the OPFFR-S, FPCA, and OPFFR methods when the number of randomly selected time points is ten.}
	\label{fig:beta-s}
\end{figure}

\begin{figure}[H]
	\centering
	\includegraphics[scale=0.6]{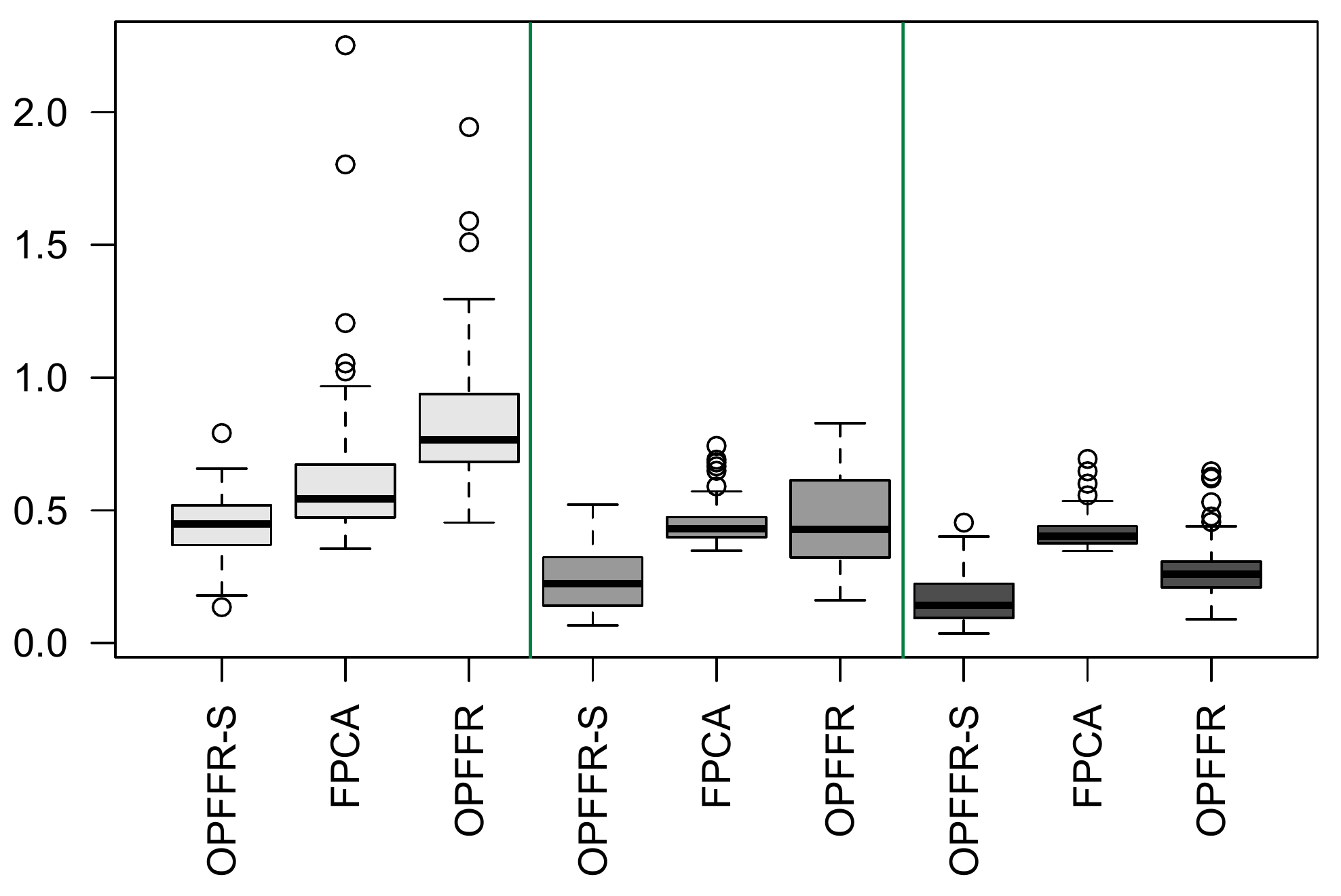}
	\caption{Boxplots of MISEs for the sparse scenario under three different sparsity levels, based on 100 simulation runs. The boxplots with different grayscale shades from left to right respectively represent the sparsity levels of 5, 10 and 15 time points per curve.}
	\label{fig:simu-re-s}
\end{figure}

Figure \ref{fig:beta-s} displayed the perspective plots of the true coefficient functions in the sparse scenario as well as their respective estimates for a single run with 10 sampled time points per curve. The  OPFFR-S method and FPCA performed well in estimating the coefficient function. The estimate recovered by the original OPFFR method was a little oversmoothed. 
In Figure \ref{fig:simu-re-s}, the performance in terms of MISEs based on 100 runs was compared. The OPFFR-S method always had the best prediction performances at all the three sparsity levels.  When the sparsity level was high (5 time points per curve), the original OPFFR method had a worse prediction performance than the FPCA. However, its prediction performance quickly picked up as the data became denser. When the sparsity level was 15 time points per curve, it actually delivered a better prediction performance than the FPCA.  Such an interesting phenomenon was referred to as the ``phase transistion'' \citep{cai2011phase,wang2016review}.

\section{Real Data Examples}
\label{sec:real}

We analyzed two real example in this section.  We showed that our method had the numerical advantage over other approaches in terms of prediction accuracy in the analysis of the Canadian weather and histone regulation data. The results in the Canadian weather data, a dense design case, and the histone regulation data, a sparse design case,  echoed with our findings in the simulation study. The smoothing parameters used in FDA for Canadian weather data were taken from the example codes in \cite{ramsay2009functional} and seven basis functions were used for the $t$ and $s$ directions respectively. In the histone regulation data we selected the smoothing parameter for FDA by a grid search on $10^{(-5\,:\,1\,:\,5)}$ and used six basis functions each for the $t$ and $s$ directions.  For the FPCA method, the `regular' parameter was set to $2$ for the Canadian weather data and $0$ for the histone regulation data. The other parameters for FDA and FPCA approaches were the same as those used in the simulation study. 

\subsection{Canadian Weather Data}

We first look at the Canadian weather data \citep{ramsay2005functional}, a benchmark data set in functional data analysis. The main goal is to predict the log daily precipitation profile based on the daily temperature profile for a geographic location in Canada. The daily temperature and precipitation data averaged over 1960 to 1994 were recorded at 35 locations in Canada. We compared OPFFR with FDA and FPCA in terms of prediction performance defined by integrated squared error (ISE) $\int_0^{365}(Y_i(t)-\eta_{\hat \beta_{-i}}(X_i, t))^2dt$, where $i = 1, \cdots, 35$ and $\hat \beta_{-i}$ was estimated by the dataset without the $i$th observation.  For the convenience of calculation, we computed $\lVert Y_i(t) -\eta_{\hat \beta_{-i}}(X_i, t) \rVert_2^2$ at a grid of values $t$ as the surrogate of ISE.
Since the findings through the coefficient function estimates were similar to those in \cite{ramsay2005functional}, we only focused on the comparison of prediction performance.  
The summary in Table ~\ref{tab:can} clearly showed the numerical advantage of the proposed OPFFR method over the FDA and FPCA methods.

\begin{table}[H]
\centering
\resizebox{1.0\textwidth}{!}{\begin{minipage}{\textwidth}
\caption{The mean, standard deviation and three quartiles of ISEs for the three approaches. The best result on each metric is in boldface.}
\label{tab:can}
\centering
\begin{tabular}{@{}cccccc@{}}
\toprule
Method          & Median & Mean & Standard Deviation & 1st Qu. & 3rd Qu. \\ \midrule
OPFFR & \textbf{21.6400} & \textbf{40.2800} & \textbf{45.7631} & \textbf{13.8000} & \textbf{36.1700} \\ \midrule
FDA & 25.9000 & 44.1600 & 56.9544 &18.7400 & 40.6100 \\ \midrule
FPCA & 30.7752 & 45.5065 & 45.7763 & 20.5031 & 52.1827 \\ \midrule
\end{tabular}
 \end{minipage}}
\end{table}

\subsection{Histone Regulation Data}

Nucleosomes, the basic units of DNA packaging in eukaryotic cells, consist of eight histone protein cores including two copies of H2A, H2B, H3, and H4. Besides the role as DNA scaffold, histones provide a complex regulatory platform for regulating gene activity \citep{wollmann2012dynamic}. Focused study of the interaction between histones and gene activity may reveal how the organisms respond to the environmental changes.
There are multiple sequence variants of histone proteins, which have some amino acid changes compared to their primary sequence, coexist in the same nucleus.
For instance, in both plants and animals, there exist three variants of H3, the H3.1, the H3.3, and the centromere-specific CENP-A (CENH3) \citep{deal2011histone}. Each variant shows distinct regulatory mechanisms over gene expression. 

In this paper, an ultra-high throughput time course study was conducted to explore the interaction mechanism between the gene activity and histone variant, H3.3, during heat stress in \textit{Arabidopsis thaliana}. In this study, the 12-day-old \textit{Arabidopsis} seedlings that had been grown at $22\,^{\circ}{\rm C}$ were subject to heat stress of $38\,^{\circ}{\rm C}$, and plants were harvested at 7 different time points within 24 hours for RNA sequencing (RNA-seq) \citep{wang2009rna} and ChIP sequencing (ChIP-seq) \citep{johnson2007genome} experiments.
We were interested in the genes responding to the heat shock, therefore 160 genes in response to heat (GO:0006951) pathway \citep{ashburner2000gene} were chosen. We selected 55 genes with the fold change above 0.5 at at least two consecutive time points in RNA-seq data.
In ChIP-seq experiments, we calculated the mean of normalized read counts by taking the average of normalized read counts over seven time points for the region of 350 base pairs (bp) in the downstream of transcription start sites (TSS) of selected 55 genes. The normalized read counts over 350 bp from ChIP-seq and the normalized fragments per kilobase of transcript per million mapped reads (FPKM) \citep{trapnell2010transcript} over seven time points from RNA-seq were used to measure the histone levels and gene expression levels respectively.

We applied the OPFFR, FDA and FPCA methods to histone regulation data in example \ref{histone}. Since the gene expression levels were sparsely observed, we also applied the OPFFR-S method to the data. The comparison of the four methods is shown in Table~\ref{tab:can2}.
In the table, the standard deviation of ISEs was the only measure that neither the OPFFR nor the OPFFR-S was the most optimal. This was caused by a few observations where all the methods failed to make a good prediction and the OPFFR methods happened to have larger ISEs. In terms of all the other measures,
the proposed OPFFR and OPFFR-S methods clearly showed the advantage in prediction accuracy again. Since the results from the OPFFR and OPFFR-S were comparable to each other, we chose to present all the following results based on the OPFFR analysis.

\begin{table}[!htbp]
	\centering
	\resizebox{1.0\textwidth}{!}{\begin{minipage}{\textwidth}
			\caption{The mean, standard deviation and three quartiles of ISEs for the four approaches. The best result on each metric is in boldface.}
			\label{tab:can2}
			\centering
			\begin{tabular}{@{}cccccc@{}}
				\toprule
				Method          & Median & Mean & Standard Deviation & 1st Qu. & 3rd Qu. \\ \midrule
				OPFFR & 1.5700 & \textbf{7.7120} & 18.9180 &  \textbf{0.5077} & \textbf{5.1900} \\ \midrule
				OPFFR-S & \textbf{1.4070} & 7.7150 & 18.6037 &  0.6972 &  5.5820 \\ \midrule
				FDA & 2.2060 & 7.9770 &  18.7004 & 0.5461 & 6.2750 \\ \midrule
				FPCA & 2.0170 & 8.4720 & \textbf{18.3978} & 0.9126 &  6.1790 \\ \midrule
			\end{tabular}
	\end{minipage}}
\end{table}

\begin{figure}[!htbp]
\centering
\includegraphics[scale=0.95]{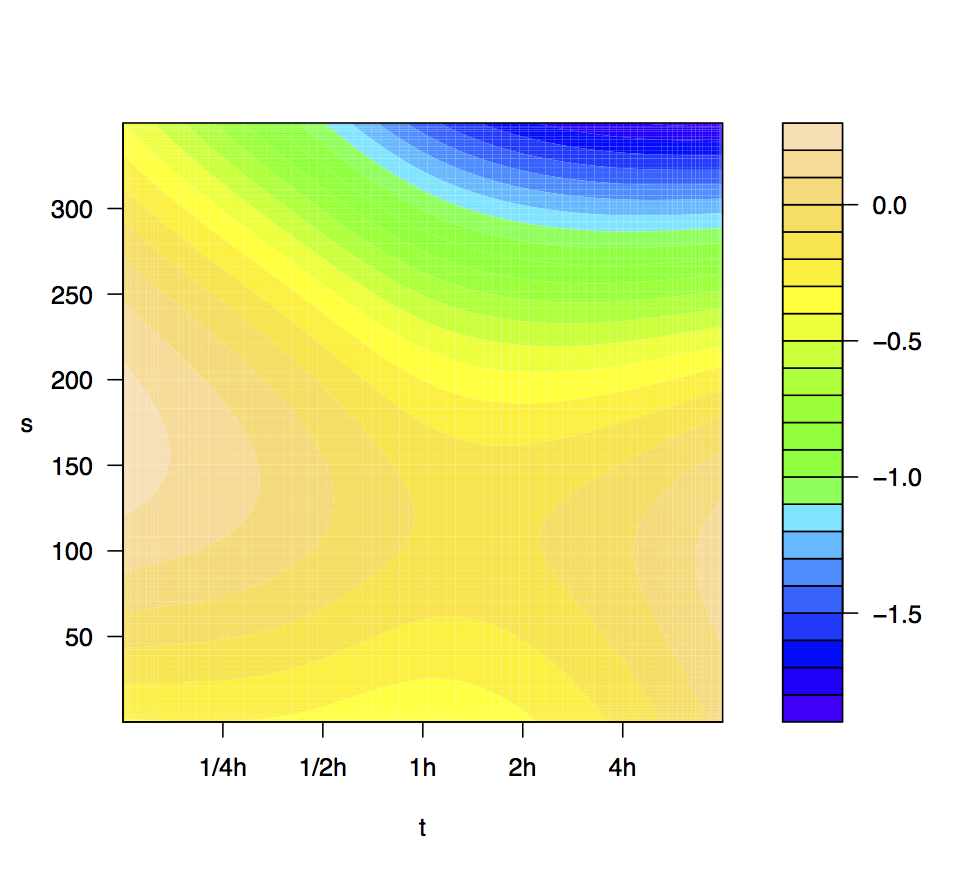}
\caption{The estimated coefficient function $\beta(t,s)$ for the histone regulation study. The y-axis label represents the positions on genomes and x-axis label represents seven time points.}
\label{fig:contour}
\end{figure}

\begin{figure}[!htbp]
\centering
\includegraphics[scale=0.45]{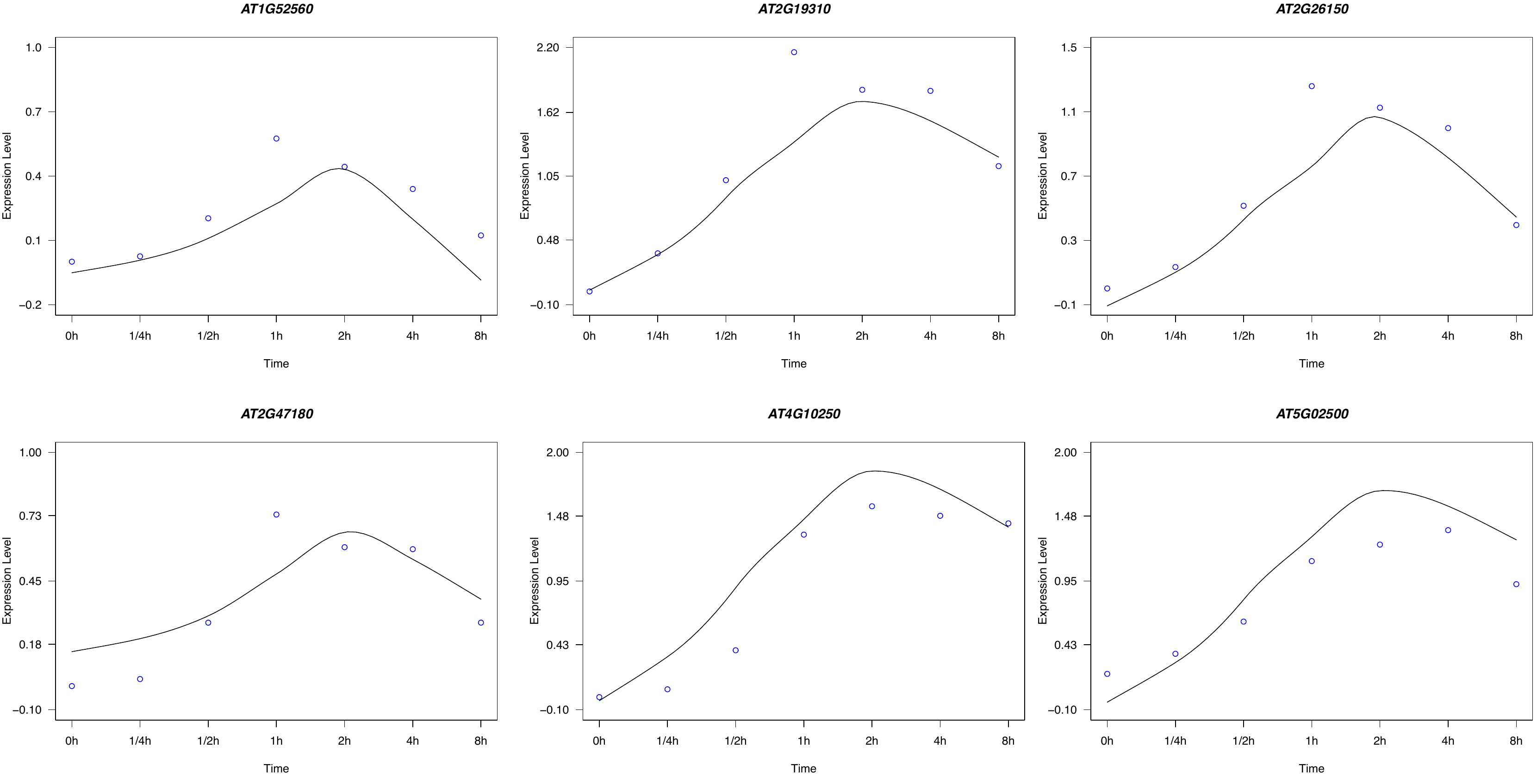}
\caption{The fitted response functions for six genes in the histone regulation study.
The y-axis stands for the normalized expression levels
and x-axis label represents seven time points. 
The curve fitted using OPFFR is in the solid line, with the data in circles.
}
\label{fig:predict}
\end{figure}

Figure~\ref{fig:contour}  is the plot of the fitted coefficient function generated from our OPFFR method. For region between 300 bp and 350 bp, there was a strong negative influence of H3.3 on genes activity from half hour to 8 hours. It indicted that the loss of H3.3 might have the biological influence on the up-regulation of heat-induced genes. This negative correlation phenomenon was also observed after 30 minutes on the region of 250 bp to 300 bp between H3.3 and gene activity.  In addition, the region from 50 bp to 150 bp had a positive effect on genes activity over time domain from 0 hour to half hour and 4 hours to 8 hours. Therefore, we provided a numerical evidence that heat-shock-induced transcription of genes in response to heat stress might be regulated via the epigenetic changes of H3.3, especially on the downstream region of TSS.
The sample plots in Figure~\ref{fig:predict} showed a nice match of the predicted gene expression curves with the observed values.
 
%

%

\section{Conclusion}
\label{sec:conc}

In this article, we have presented a new analysis tool for modeling the relationship of a functional response against a functional predictor.
The proposed method is more flexible and generally delivers a better numerical performance than the FPCA approach since it does not have the restrictive structural dependence assumption on the coefficient function.
When compared with the penalized B-splines method, the proposed method has the theoretical advantage of possessing the optimal rate for mean prediction as well as some numerical advantage as shown in the numerical studies. Moreover, the Representer Theorem guarantees an exact solution to the penalized least squares, a property that is not shared by the existing penalized function-on-function regression models. The application of our method to a histone regulation study provided numerical evidence that the changes in H3.3 might regulate some genes through transcription regulations. Although such a finding sheds light on the relationship between histone variant H3.3 and gene activity, the details of the regulation process are still unknown and merit further investigations. For instance, we may investigate how the H3.3 organizes the chromatins to up-regulate those active genes. Such investigations would call for more collaborations between statisticians and biologists.  	

When the regression model has a scalar response against one or more functional predictors, methods other than the roughness penalty approach are available to overcome the inefficient basis representation drawback in the FPCA method. For example, \cite{delaigle2012methodology} 
	considered a partial least squares (PLS) based approach.  \cite{ferre2003functional} and \cite{yao2015effective} translated the idea of sufficient dimension reduction (SDR) into the setting of functional regression models.  Intuitively, these methods might be more efficient in their selection of the principal component basis functions since they incorporate the response information into consideration. However, our experiments with a functional response version of the functional PLS \citep{Preda2005149}, not shown here due to space limit, did not look so promising. Therefore, further investigation in this direction is surely needed.

\newpage

\bibliographystyle{agsm}
\bibliography{main}
\end{document}